%% file: hicss.tex
\documentclass[10pt]{article}
\input{hicss-packages.tex}

\title{Representative Dataset Generation Framework for AI-based Failure Analysis during real-time Validation of Automotive Software Systems }

 \author{Mohammad Abboush \\
 Technische Universität Clausthal,\\ Institute for Software and Systems \\Engineering \\
  {\underline{ mohammad.abboush@tu-clausthal.de}} 
    \And
  Christoph Knieke \\
 Technische Universität Clausthal, \\Institute for Software and Systems \\Engineering \\
  {\underline{ christoph.knieke@tu-clausthal.de} } 
   \And
  Andreas Rausch \\
 Technische Universität Clausthal,\\ Institute for Software and Systems \\Engineering \\
  {\underline{ andreas.rausch@tu-clausthal.de} }   }
  
\date{}
\begin{document}
\maketitle
\begin{abstract}

Recently, thanks to its ability of extracting knowledge from historical datasets, the data-driven approach has been widely used in various phases of the system development life cycle. During real-time system validation, remarkable achievements have been accomplished in developing an intelligent failure analysis based on historical data. However, despite its superiority over other conventional approaches, e.g., model-based and signal-based, the availability of representative datasets persists as a major challenge. Thus, for different engineering applications, new solutions to generate representative faulty data in different forms should be explored. Therefore, in this study, a novel approach based on Hardware-in-the-Loop (HIL) simulation and real-time Fault Injection (FI) method is proposed to generate and collect data  samples under single and simultaneous faults for Machine Learning (ML) applications during system validation phases. The developed framework can generate not only sequential data, but also textual data including fault logs. The results show the applicability of the proposed framework in simulating and capturing the system behaviour under faults within the system components. 
\end{abstract}

\subsubsection*{Keywords:}

HIL testing, fault injection, Automotive systems development, failure analysis, machine learning.

\section{Introduction}

In accordance with model-based design and the V-model development approach, the System Under Test (SUT) is tested at each level of the development process. In other words, the testing and validation process are performed for different states of the SUT, i.e., the executable model, the model code and the implemented code on the host and target machine (\cite{garousi2018testing}). Several platforms have been leveraged to carry out the testing and validation activities, which are known as X-in-the-Loop (\cite{shokry2009model}). As such, Hardware-in-the-Loop (HIL) real-time simulation is recommended by ISO 26262 as a safe, flexible, reliable, and effective platform (\cite{himmler2012hardware}). Recently, HIL has played a vital role in the verification and validation of automotive embedded control systems under time constraints. Substituting the real hardware elements with fidelity simulated system provides not only avoidance of potential risks but also reductions in testing costs in various applications e.g., electric drives, power electronics, power grids, railways and automotive (\cite{mihalivc2022hardware}). HIL-based system validation is currently a hot topic in academia and industry due to its ability to provide efficient, fast, and realistic simulations in real-time with high accuracy. On top of that, by serving as a digital test drive platform, HIL has contributed to overcome the limitations of real test drives in terms of time, cost, effort and risk for the tester (\cite{chen2018autonomous}). However, at system integration testing level, an enormous amount of datasets from heterogeneous components and subsystems are generated as a result of tests execution (\cite{jordan2020time}). Besides, despite test automation of test cases (TCs), many failed TCs are documented in the test report as pass/fail (\cite{vermeulen2008functional}). Therefore, analysing the vast amount of test records based on traditional approaches is costly, difficult, and time-consuming (\cite{nair2017data}). This is why an intelligent system capable of analysing the test results of HIL in an efficient way without domain knowledge is necessary.

In the last decade, the availability of sufficient computational resources paved the way for the widespread application of data-driven approach to tackle the problem of Fault Detection and Diagnosis (FDD). Central to this approach is the idea of extracting knowledge from historical data and constructing non-linear relationships between input and output classes to discover hidden patterns. Compared to other FDD approaches, neither does it require expert knowledge nor a precise mathematical model. Hence, the data-driven approach has attracted the attention of researchers, and the development of methods has increased rapidly in various technical fields. Deep Learning (DL) and Machine Learning (ML)-based methods, as a subset of the data-driven approach, are widely used in various phases of the software development life cycle, i.e., software requirements, software architecture and design, software implementation, software quality and analysis, and software maintenance (\cite{shafiq2021literature}). In the testing and analysis phase, for example, DL and ML extend not only to the generation and selection of TCs, but also to the detection and analysis of defects and anomalies. Moreover, the methods have been used to construct a model capable of determining the causes of failed TCs based on historical logs (\cite{zhang2011automated}). 
However, despite the remarkable accomplishments in different domains, one of the major challenges impeding DL and ML methods is the availability of datasets  (\cite{badihi2022comprehensive,neupane2020bearing}). Resultantly, non-availability of datasets imposes constraints on the development process of DL-based FDD in real-world applications. For that reason, establishing representative datasets, such as document logs and signal data, should be further explored.

As previously mentioned, the core element of developing FDD model is the dataset. In principle, obtaining fault-free data can be realised by logging the system behaviour in fault-free mode (\cite{zhang2020deep}). However, acquiring representative data as a reaction to the system behaviour in case of a fault occurrence is a complicating factor with regards to difficulty and cost (\cite{zhang2022intelligent}). In automotive domain, due to confidentiality of test data, it is very rare to have public real-world data containing faulty behaviour (\cite{theissler2021predictive}). Besides, even in the best case with limited data, the ratio of faulty to healthy is small and unstable, which in turn leads to the problem of imbalanced data. A reason for this is the difficulty in simulating the open set of failure modes under complex working conditions (\cite{kukkala2018advanced}). Besides, capturing hardware faults is not feasible in real-world applications (\cite{fernandes2022machine}).

In order to train a robust intelligent FDD, the following requirements on the dataset should be ensured:
1. a high quality dataset with a sufficient number of samples, including healthy and faulty operating conditions, 2. a high degree of fault classes coverage, 3. consideration of the single and simultaneous faults occurrence, 4. consideration of real-time constraints on system behaviour under normal and faulty conditions. 5. offering different types of datasets, i.e., time series and text data.

Therefore, tackling the problem of representative dataset is still an open concern, and collecting faulty dataset that meet the above requirements should be further explored. In this paper, we attempt to bridge this gap by proposing a novel framework for generating and collecting representative data including healthy and faulty samples during Automotive Software Systems (ASSs) development. To demonstrate the benefits and applicability of the proposed framework, a high-fidelity simulation of a gasoline engine system is used as a case study. Vehicle dynamics, environment, driver and powertrain models have also been considered to capture the detailed characteristics of the vehicle during data acquisition.  The main contributions of the proposed framework can be summarised as follows:
\begin{itemize}
\setlength\itemsep{0em}
\item We propose a novel framework capable of generating real-time faulty datasets, taking into account the time constraints of the system behaviour.
\item The framework can cover major types of random hardware sensor and actuator faults in ASSs.
\item Not only the single faults but also the simultaneous occurrence of faults can be simulated, providing a high coverage.
\item Real-time HIL simulation and high fidelity simulation models are employed providing high quality real-time coverage of the collected data.
\item Finally, our proposed framework provides two types of representative datasets to be collected, i.e., time series and text log data. 
\end{itemize}
The rest of the paper is structured as follows:
Related work and other contributions are presented in Section 2. Section 3 introduces the proposed approach, highlighting the key stages of dataset collection and preparation. The implementation steps and the case study for validation are described in Section 4 . Section 5 summarises the results and findings. Finally, Section 6 presents the conclusion and future work.

\section{Related work}

To tackle the problem of faulty data availability, several solutions have been proposed in the literature. Some of them have used online public datasets, while others had to generate the data depending on the real prototype or simulated system. For example, (\cite{rengasamy2020deep}) have used a standard dataset of a gas turbine engine provided by NASA. The mentioned data is used to develop a DL-based model for prediction and diagnosis tasks. Similarly, based on Audi's industrial dataset, (\cite{safavi2021multi}) developed an intelligent model relying on DL methods for fault detection, isolation, identification and prediction. The target system in the work is automotive autonomous driving. Despite the advantage of employing published standard data, the fault classes are limited depending on the source, lacking the flexibility to extend the data under the same operating conditions. A static injection of the faulty sample into the data, based on normal distributions with one standard deviation, has been used as an alternative solution. Employing a real prototype has been an alternative solution for other researchers. For example, aiming at developing a robust FDD for vehicle engines, (\cite{jung2020data}) employed a real engine to capture the system behaviour under normal and abnormal conditions. Since the data collected exactly matched the real industrial application data, the developed model showed a high degree of applicability and robustness to the environmental conditions, e.g., noise. However, the major limitation of the applied technique is the potential risk and damage to the target physical system in case of Fault Injection (FI). In the same respect, but for different applications, (\cite{bafroui2014application}) and (\cite{gharavian2013comparison}) have proposed a FDD model for automotive gearboxes. The developed model was structured based on data collected on an experimental test bench in the laboratory. Some other studies have been conducted based on data collected from real vehicle prototypes to develop an intelligent solution for the detection of unknown faults in the test drive records (\cite{theissler2017detecting}). However, beside the high cost, this type of experimentation could pose a risk not only to the vehicle, but also to the tester in case of a critical situation. Moreover, some types of faults cannot be explored, in which the whole physical system can be damaged.

To overcome the limitations of using real physical systems, a simulation platform has been used to generate representative datasets.  Based on the modelling and simulation techniques, many researchers have proposed a solution to address the problem of availability data (\cite{al2016remaining}). For example, (\cite{tagawa2015structured}) have employed the FI method to generate the faulty data based on the simulated system in the MATLAB/Simulink environment. In this work, four driving scenarios were used to generate faulty data along with normal data. Similarly, (\cite{biddle2021novel}) approached the problem of faulty data availability by injecting five different types of faults into the simulated system, namely erratic, hardover, drift, spike and stuck fault. To this end, the failure modes were modelled in the simulation environment, i.e. MATLAB/IPG CarMaker co-simulation, and then the faults were artificially injected according to the fault parameters. Consequently, and based on the collected data, ML-based architecture for multi-fault in multi-sensor FDD was constructed. However, despite the ability to reproduce the test under critical conditions with a high degree of confidence, real-time constraints remain ignored. The effect of this is that the application of the target model in the real world becomes increasingly restricted. Not only that, the current developed FDD models are validated based on simulation data generated by a simulation platform. By doing so, the real industrial conditions, such as influences, noise and uncertainties, are ignored (\cite{yin2022review}).
Lately, with the aim of overcoming the limitations of pure simulation, HIL real-time simulation has been introduced as a solution for generating data covering various critical conditions (\cite{gonzalez2021data}). Employing Dataset generated from HIL, (\cite{abboush2022intelligent}) have proposed FDC model based on hybrid DL techniques to be used during the development phases of ASSs, i.e. integration testing. The basis of the developed model is the faulty data collected by programmatically injecting nine different single sensor faults into the target system in real-time without changing the model. As an  extension of the mentioned work, in this study, the simultaneous occurrence of the faults are  considered, including transient and permanent occurrence. Besides, the textual nature of logs from test execution results are collected for Root Causes Analysis (RCA) problem.

\section{Methodology}
\begin{figure*}[thb]
    \centering
	\includegraphics[trim={0.5cm 0.5cm 0.5cm 0.5cm}, clip,width=1\linewidth]{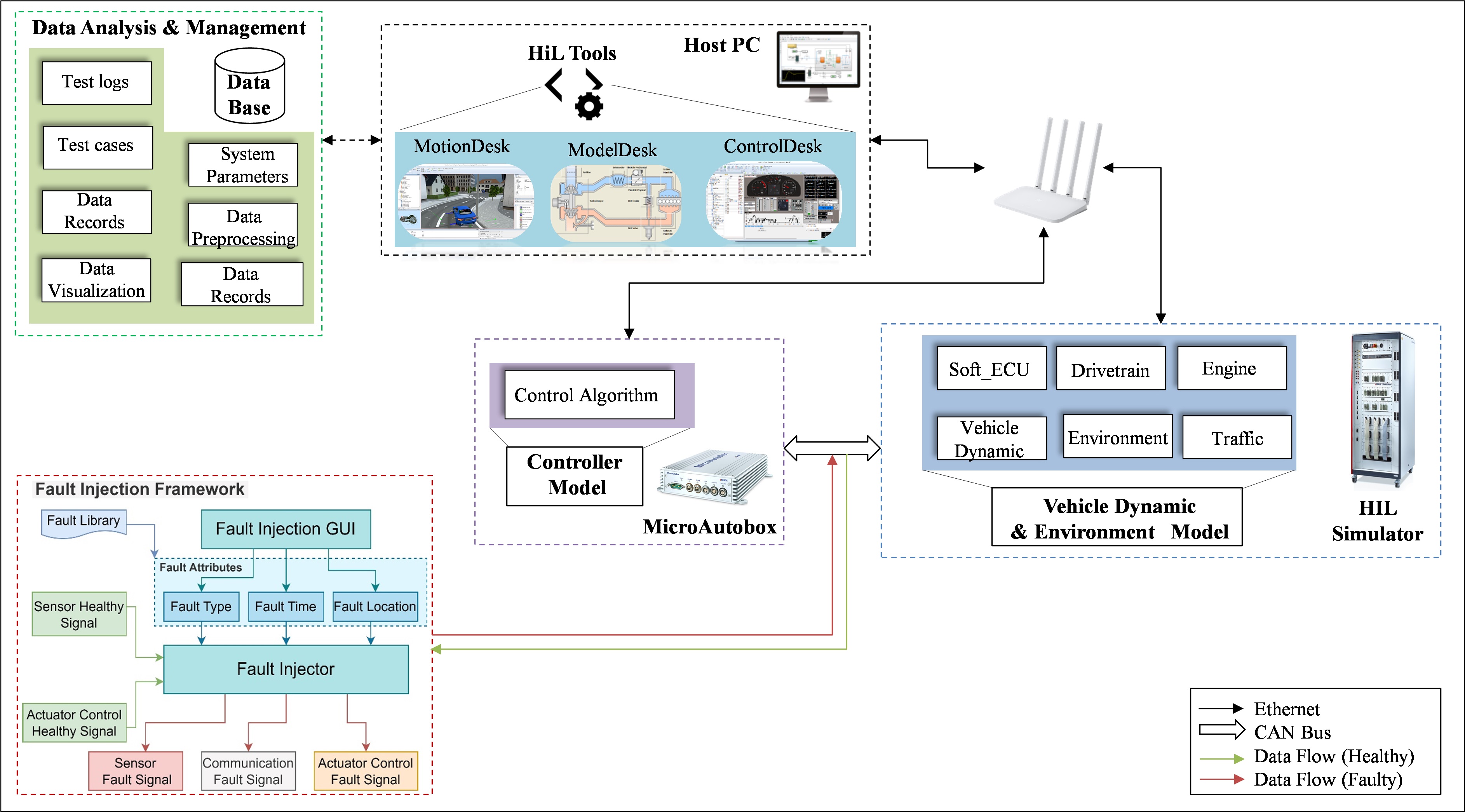}
	\caption{Proposed framework for representative dataset generation.}
	\label{fig: sample-figure}       
\end{figure*}
In this section, the proposed framework is presented, including real-time HIL simulation, data analysis and management, and FI, as shown in Figure 1. 

The HIL system is the core of the framework, in which the complex target system is simulated and executed in real-time. It consists of two main parts, namely the HIL simulator and the target prototype (controller). HIL simulator is responsible for executing the controlled system (plant) in real-time. To be precise, in our case the controlled system is the entire vehicle model. Engine model, dynamic vehicle model, environmental model, traffic model and powertrain model are the main subsystems of the selected system.  Noteworthy, the SoftECU, i.e., virtual ECU, model is also included in the controlled system model. By doing so, the whole vehicle model including the controller model can be executed in the HIL simulator, which act as a Rapid Control Prototype (RCP), known as offline mode. On the other hand, Microautobox II is the target machine of the SUT where the ECU model is deployed and executed. Microautobox II is considered as RCP and acts as the real ECU. Both parts of the HIL system are connected via the CAN bus. To establish the connection, the signal interface is modelled in the simulation environment on both sides, i.e. in the ECU model and in the plant model. The generated code of the mentioned models is respectively injected from the host PC via Ethernet into the control unit and the HIL simulator.
On the host PC, the configurations and experiments are carried out. For this purpose, four software tools from dSPACE are used, namely MotionDesk, ConfigurationDesk, ModelDesk and ControlDesk (\cite{dSPACE}). In addition to the parameterisation of the system model, ModelDesk is used to design the test drive scenarios and the TCs. Thanks to the model-based design approach, once the model is configured, the code of the both models, i.e., SUT and plant, is automatically generated and deployed using ConfigurationDesk. To represent the driving environment and dynamic traffic as a 3D visualisation, MotionDesk is employed. Finally, the instrumentation, measurements, dataset acquisition and control of the experiments during real-time execution are performed using ControlDesk. The mentioned tool enables also switching between the offline execution mode (SoftECU) and the online execution model (Real ECU).

FI process takes place at the signal interface between the control unit and the HIL simulator during real-time execution. By doing so, the execution of the system model of the ECU and the plant can be ensured in real-time as a black box without modification. Furthermore, the injection process is executed programmatically without extending the target system model with additional components. The input of the FI framework is healthy data representing the system behaviour under normal conditions. The target signals are then manipulated according to the fault attributes configured by the user. There are three attributes in the fault injector that should be configured before injection, i.e., fault type, fault location and FI time. Fault type covers most sensor and actuator malfunctions such as gain, offset, hard over, stuck-at, delay, noise, packet loss, drift and spike fault. The target sensor and actuator signal is identified as a location for faults to be injected. Duration of the FI, and the time at which the fault is injected, are specified according to the driving cycle or standard system behaviour.
Finally, the data analysis and management is carried out using host PC. Here, the system requirements, the TCs specifications and the scenarios description are documented. Besides, in this phase the collected healthy and faulty datasets are stored as a result of the test execution. Two types of data are collected, sequential signal data and textual log data. This allows the collected data to be used for various tasks and applications, e.g. fault detection, diagnosis, prognosis and RCA of failed TCs. In addition to data management, pre-processing of the data takes place so that outliers and redundant data are removed. Beyond that, the data is converted into a suitable format, e.g. CSV, MAT or XML, to be ready for the ML/DL model development process. Finally, the pre-processed data is visualised on the host PC to get a deeper understanding of the patterns and features before training the DL/ML model.

\section{Case study and Experimental Implementation}

To validate the applicability of the proposed framework, this section presents the selected case study, highlighting the system architecture and implementation steps.

\subsection{Case study}

As a case study from the automotive domain, gasoline engine system from dSPCAE (\cite{ASMdSPACE}) was used to validate the proposed approach. The architecture of the system is shown in Figure 2. What can be seen is that the various systems and subsystems have been modelled to accurately reflect real-world behaviour. MATLAB/Simulink was selected as the graphical simulation and modelling environment to model the dynamic system. It has significant features in designing, analysing and testing the dynamic system using a model-based approach. Besides, it offers the possibility to test the SUT in a simulated environment and automatically generate the code before deployment, supporting real-time simulation. However, despite the realistic simulation of the selected system, it does not enable drive with lateral control. To overcome this limitation, in our case , a complex gasoline engine model was integrated into the dynamic system model of the vehicle. Thereby, lateral and longitudinal driving with manual and automatic transmission can be simulated. The major subsystem models that structure the gasoline engine are exhaust, fuel, air path, cooler and the piston engine system. Various components have been modelled and interconnected in a block diagram environment so that the comprehensive functions of the gasoline engine are simulated. Additionally, the powertrain, driver and environmental model are included to capture the overall dynamic characteristics of the vehicle. The gasoline engine is controlled by the ECU in two modes, i.e. offline and online. In offline control, the soft ECU is internally connected to the engine, whereas a separate ECU model in the RCP is used as the real ECU to enable online control. Finally, the interface between the real ECU model and the gasoline engine model is realised using a real-time interface CAN multimessage blockset (RTICANMM) wherein the fault injections are configured and modelled. Thanks to the feature of RTICANMM, the system signals can be accessed and manipulated programmatically without changing the original system model.

\begin{figure}[thb]
    \centering
	\includegraphics[trim={0.1cm 0.1cm 0.1cm 0.1cm}, clip,width=1\linewidth]{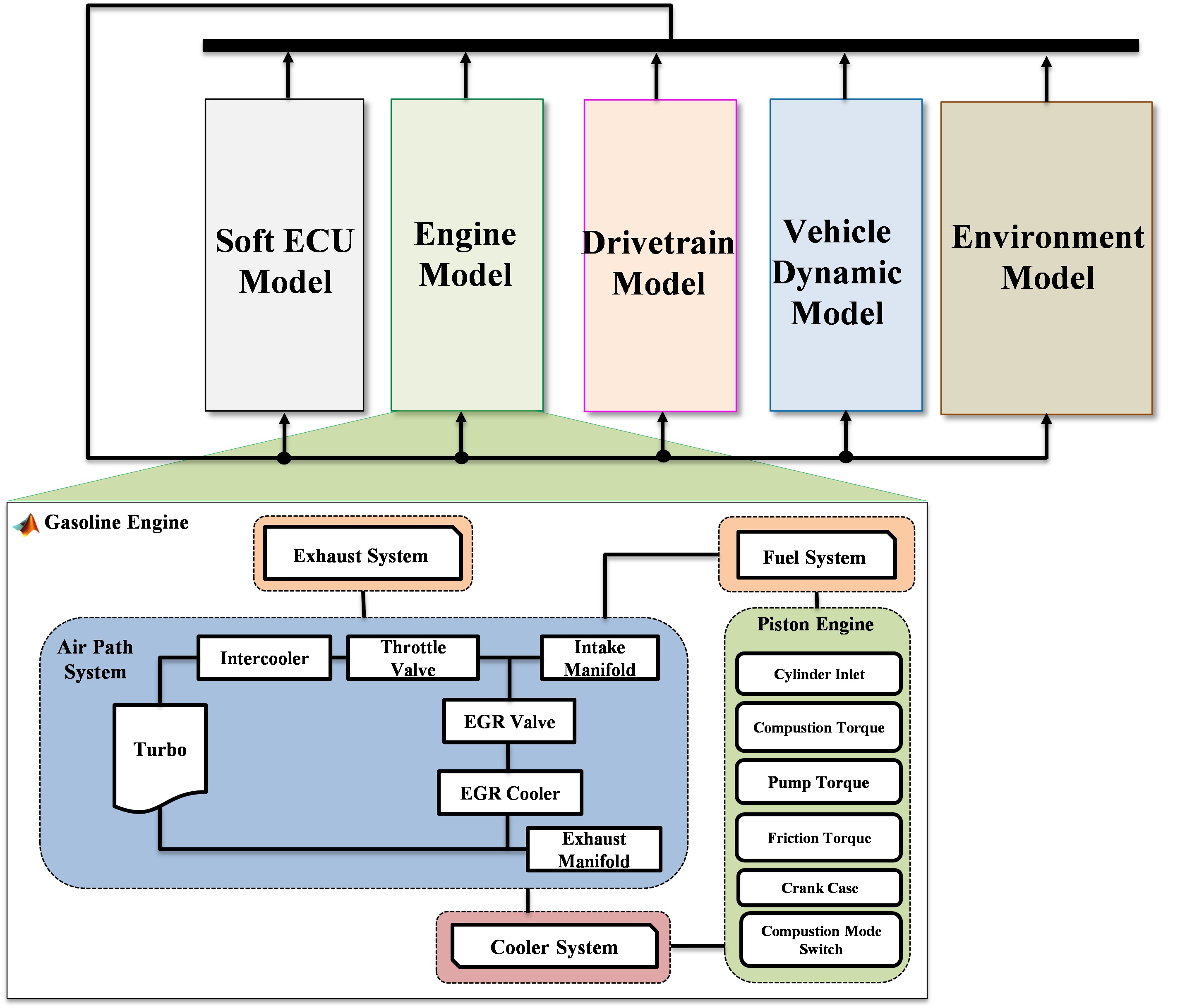}
	\caption{System architecture of the case study.}
	\label{fig: sample-figure}       
\end{figure}

\subsection{Experimental setup}

There are three levels of configurations in our case study, namely system model configurations, experiments configurations and FI configurations. 
At the first level, dSPACE's software tools, i.e. ModelDesk and ConfigurationDesk, are used to specify the parameters and variables of the models based on the GUI. Additionally, driving scenarios and TCs are designed based on the specifications. In our case, two driving scenarios are selected, namely Highway and Ftp 75. By doing so, driving scenarios on the highway, on non-urban/open roads and in urban traffic can be covered. Once the system and test configurations are specified, the code can be automatically generated and deployed to the target machines using ConfigurationDesk. Two target machines are employed in our study, Microautobix II for the ECU system and the HIL SCALEXIO simulator for the entire controlled vehicle system. Finally, the 3D visualisation of the driving environment, i.e. the defined scenarios, is modelled with MotionDesk. 
On the sec level, the configurations of the fault injector are carried out. To be specific, the FI attributes, fault location, fault type and FI time, are configured. As already mentioned, all system signals, sensors and actuators, can be accessed via the CAN bus interface. Therefore, the degree of coverage of our developed framework is high compared to the traditional approach. Some examples of sensors signals accessed are crank angle, battery voltage, engine speed, accelerator pedal position (APP), ignition and starter request, EGR mass flow, intake and exhaust manifold pressure, fuel pressure, coolant temperature and railbar sensor. The actuators involved are the throttle valve, the fuel metering unit, the pressure control valve and the ignition angle adjuster. Due to the significant effect of a fault in the APP and engine speed sensor on vehicle behaviour, we selected these locations for FI in our study. On the other hand, nine different types of the single faults and 15 combinations have been identified to demonstrate the coverage of fault data generation. By this way, not only the single fault-based dataset but also the concurrent faults-based dataset can be collected. Finally, when to inject the fault and the duration of the injection are determined based on the selected driving cycle. FI can be permanent until the faulty component is treated, or it can be temporary occurring frequently for a set duration. By injecting permanent and transient faults, both balanced and imbalanced datasets can be acquired for developing a reliable FDD model for real-world applications. The detailed configuration of the FI framework for single and simultaneous faults is shown in (\cite{abboush2022hardware}). 
The experimental setup takes place at the last level, where the measurements, recording and instrumentation are carried out. The sampling time of the signal measurements is set to 0.01 sec in this study. At the system level, 6 variables of the target system are selected to represent the recorded system behaviour under normal and faulty conditions. These system variables are throttle position, engine temperature, mean effective engine torque, engine speed, intake manifold pressure, rail pressure and vehicle speed, as illustrated in Figure 3. 

\begin{figure}[thb]
    \centering
	\includegraphics[trim={0.1cm 0.1cm 0.1cm 0.1cm}, clip,width=1\linewidth]{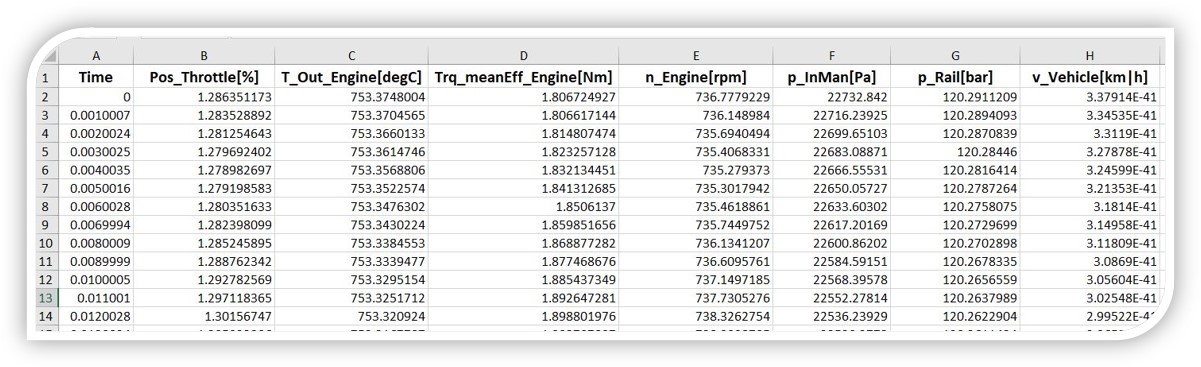}
	\caption{Generated Dataset}
	\label{fig: sample-figure}       
\end{figure}

With the aid of ControlDesk, which can be used to configure and control the run-time experiments, the system is executed under the various conditions. Once the real-time execution is completed, system response to the faults is captured as time-series data in a CSV file containing system variables and data samples. 
On the other hand, TCs designed in ModelDesk are executed on the target SUT to generate the documented error logs as text data. The TCs should contain the data inputs and the expected outputs. According to the actual system behaviour, the defined expected output is then compared with the measured output. According to the test execution, the results can be either pass or fail. Thanks to the ASM test function of ModelDesk, the TCs are automatically executed and evaluated. Note that the generated report contains key information about the TCs specifications as well as the execution results (passed or failed). Besides, the report contains a failure log describing the occurred failure. With the logs of the failed TCs, an intelligent model for RCA is developed based on natural language processing and machine learning methods.

\section{Result and Discussion }

In this section, the results of the data collection for the different FI configurations are presented. Specifically, the generated data is described, highlighting the effects of single and concurrent faults on system behaviour. Besides, the log data of failed TCs generated in the test report are presented.

In the real-time, both the SUT and the controlled system are executed under normal and abnormal conditions according to the aforementioned system and FI configurations. A total of 40 experiments were conducted with single and simultaneous faults in two driving scenarios. In particular, two files as healthy data from highway and ftp 75 scenarios, 18 files represent the single fault types and 30 files as simultaneous fault with 15 combinations have been collected. As a result, 50 CSV files have been obtained as HIL testing records. 

Focusing on packet loss and stuck-at as an example of fault types, the system behaviour under the injected fault can be captured at the system level on vehicle and engine speed signals. In Figure 4, the stuck-at fault has been injected into the RPM sensor after 10 sec of driving cycle. The engine's behaviour deviates significantly from 750 rpm in the normal range to 2300 rpm as soon as the fault is activated, which in turn leads to an increase in energy consumption. Until the 180 sec, the control unit is unable to cope with the fault's occurrence. However, after 180 sec, the system tries to overcome the fault and behave according to the desired behaviour. Then the deviation and fluctuations  take place again from 210 sec and 290 sec, respectively. The reason behind this change is the driving variation and conditions between the acceleration mode, deceleration mode and steady state mode. This means, in other words, that the control strategy cannot properly mitigate the fault when the vehicle is accelerating and decelerating sharply, as seen in Figure 5. 

\begin{figure}[thb]
    \centering
	\includegraphics[trim={0.2cm 0.2cm 0.2cm 0.2cm}, clip,width=1\linewidth]{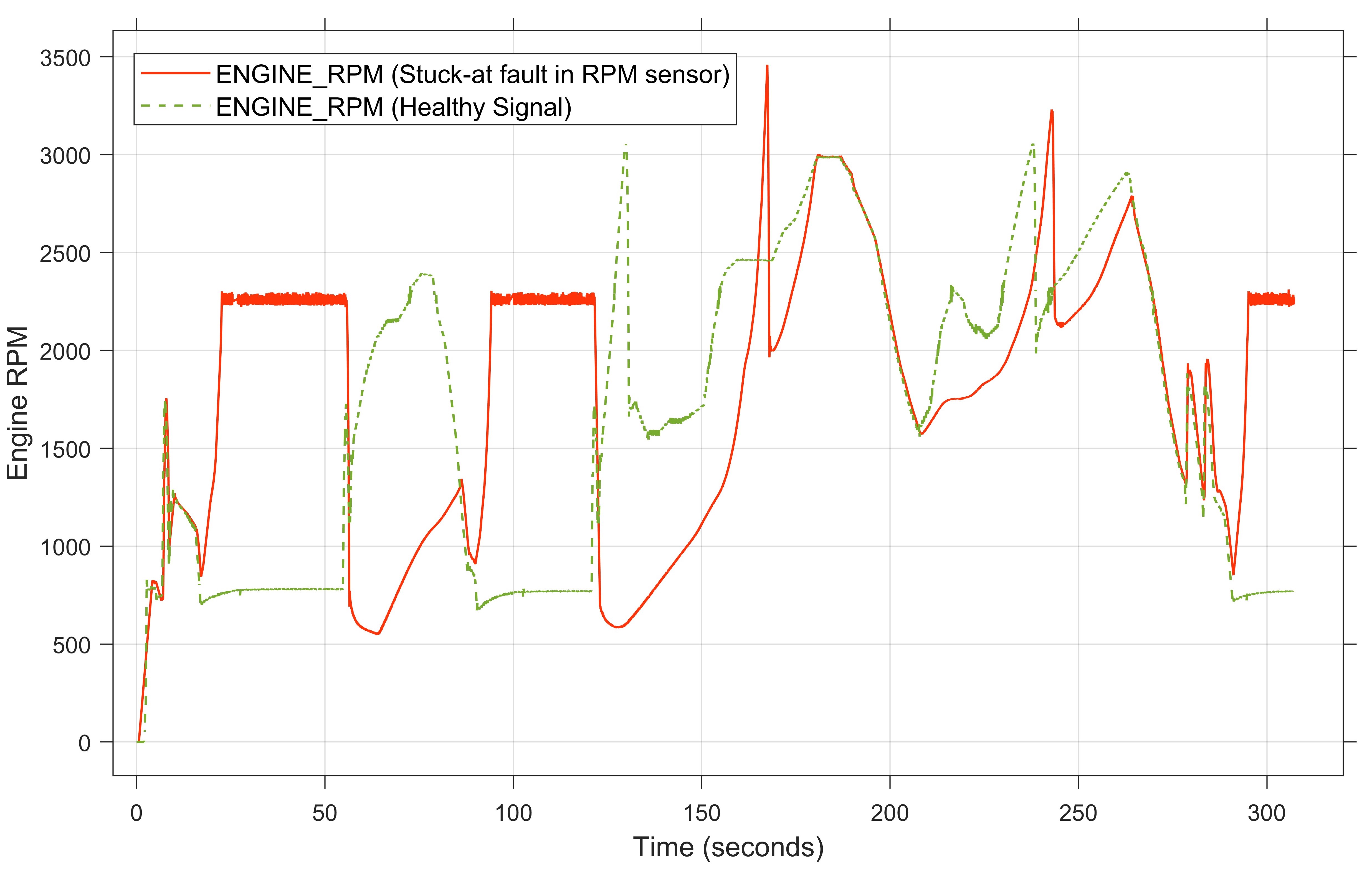}
	\caption{Engine system behaviour under Stuck-at fault}
	\label{fig: sample-figure}       
\end{figure}

\begin{figure}[thb]
    \centering
	\includegraphics[trim={0.2cm 0.2cm 0.2cm 0.2cm}, clip,width=1\linewidth]{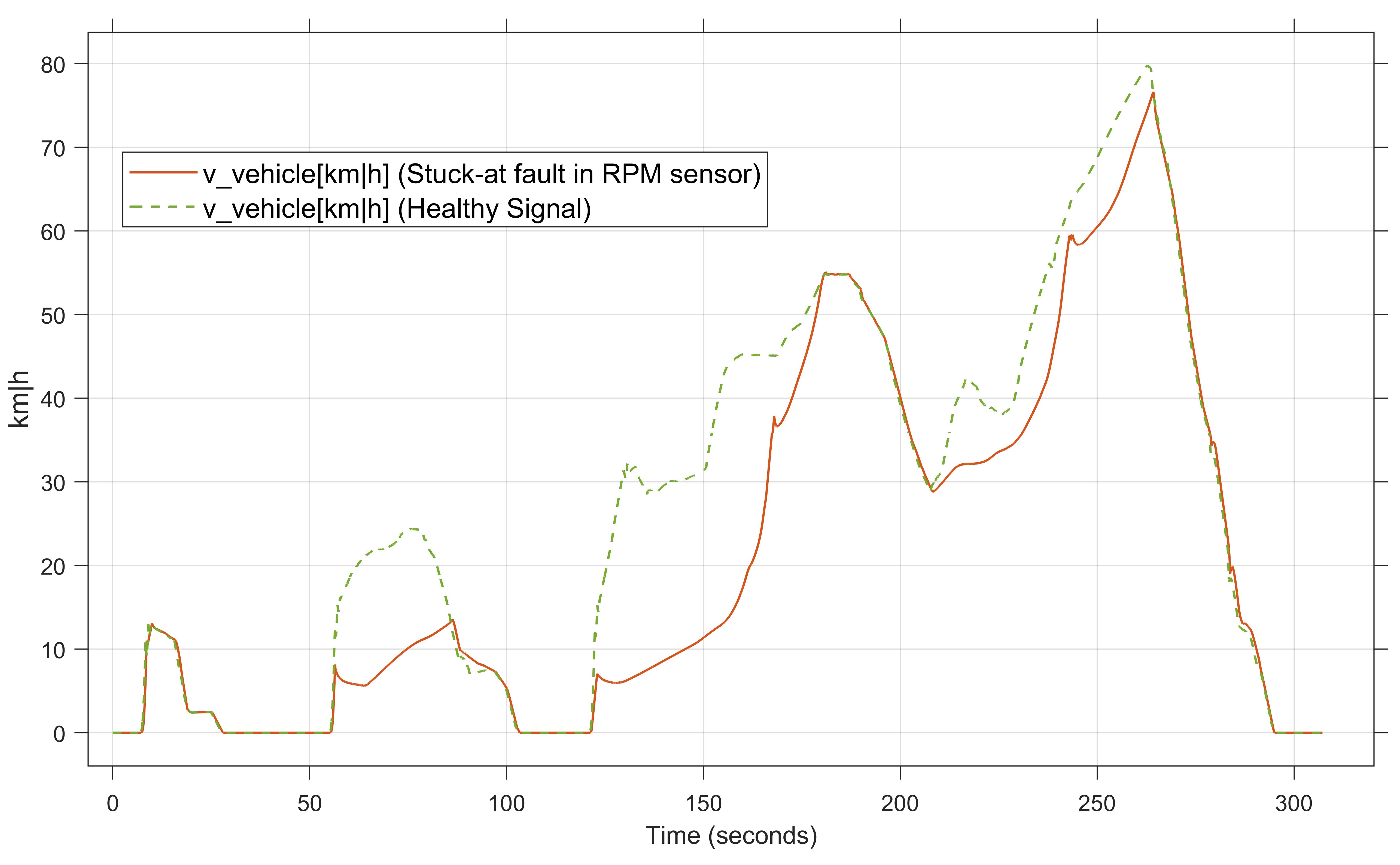}
	\caption{Vehicle system behaviour under Stuck-at fault}
	\label{fig: sample-figure}       
\end{figure}

On the other hand, by injecting the packet loss into the APP sensor during the driving cycle, the effect of the fault cannot be directly observed on the vehicle speed and engine speed, as presented in Figure 6. However, from 75 sec onwards, significant fluctuations can be observed. At the system level, the behaviour of the vehicle and the engine deviates from the standard in the form of severe jerking. At this moment, the ECU tries to overcome the signal losses for a certain period of time. That is due to the fact that the behaviour remains in the state of fluctuation when the loss time is less than 3 sec.

\begin{figure}[thb]
    \centering
	\includegraphics[trim={0.2cm 0.2cm 0.2cm 0.2cm}, clip,width=1\linewidth]{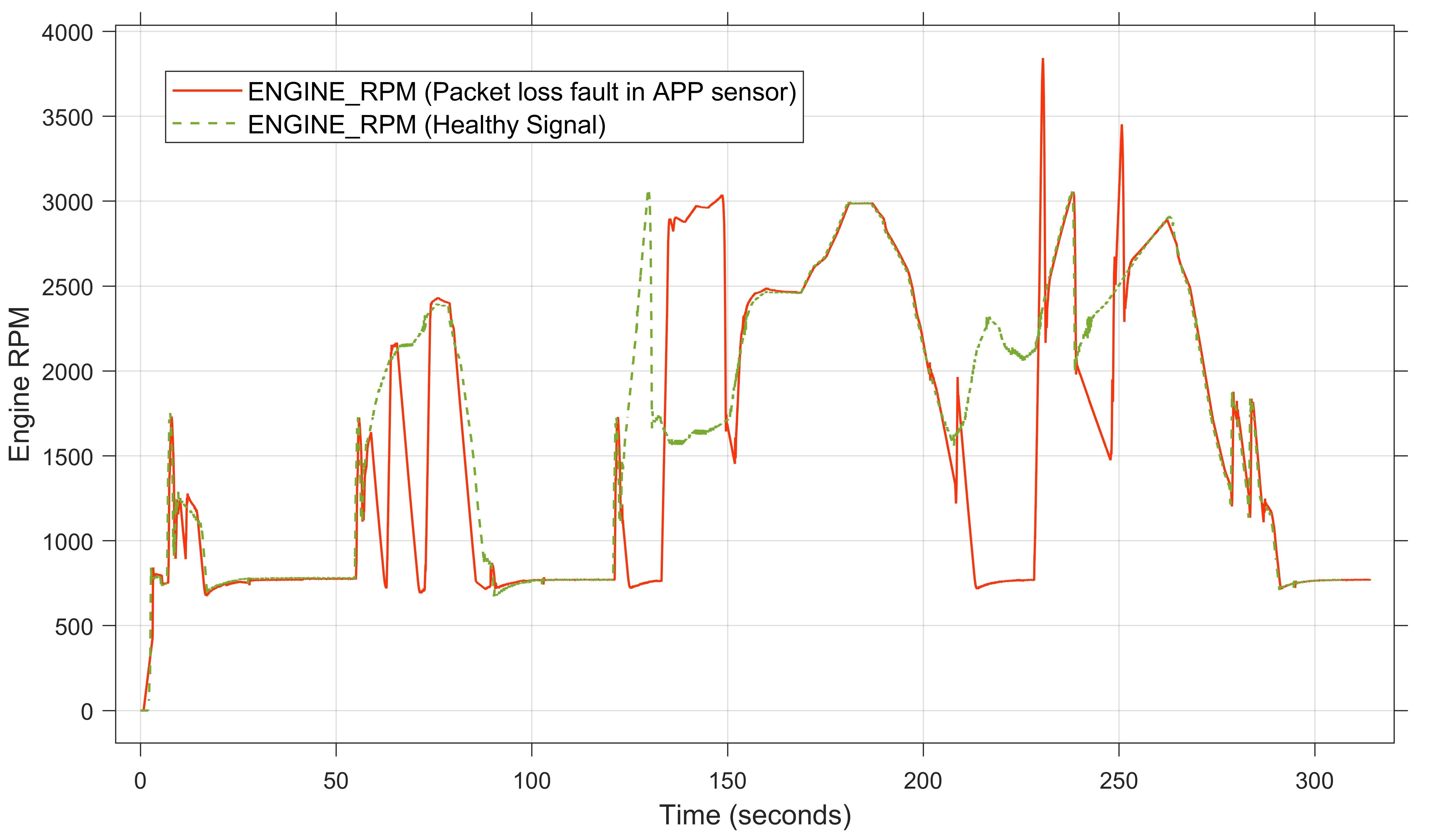}
	\caption{Engine system behaviour under packet loss fault}
	\label{fig: sample-figure}       
\end{figure}

According to the ftp 75 scenario, what can be seen in Figure 7 and Figure 8 is the effect of injecting a single fault into the APP and RPM sensor, respectively. Specifically, APP sensor signal is manipulated by the stuck-at value, and the effect of the fault is directly observed at the system level as a constant value, i.e. 1160 rpm. On the other hand, the effect of the delay of the RPM sensor signal causes fluctuations of the engine speed between 265-307 sec, as shown in Figure 8. During this period, the ECU cannot perfectly perform the desired behaviour with increasing delay of the received signal, especially when the driving mode is changed.

In the case of the simultaneous faults, two different fault types are injected simultaneously into two different locations. For example, while the stuck-at fault is injected into the APP sensor, the delay fault is simultaneously injected into the engine speed sensor. To put it in another way, two or more factors have contributed to creating a novel pattern in the signals as an effect of the simultaneous FI. Figure 9 illustrates the effect of the simultaneous faults on the vehicle system behaviour, i.e., vehicle speed, between 170-330 sec. The vehicle speed closely follows the desired behaviour up to 170 sec, at which point the faults are activated. As a result of the large fluctuations shown in Figure 10, the vehicle is unable to accelerate at 190 sec, at which time the speed drops to 18 km/h at 216 sec, causing a risk of failure. However, the system returns to the acceptable state with minor deviations. 

In spite of the fact that the collected data represent comprehensive characteristics of the system behaviour, the generated csv files contain redundant or useless information. For example, each file contains information about the recoding process as well as the data samples. Hence, the collected data should be preprocessed to remove outliers and redundant information. Three steps are applied to the data in the pre-processing phase, i.e. usefulness information and outlier removal, data scaling and normalisation, and data splitting.  After converting the csv files into execl format, the training variables are selected and the non-useful data is removed. By doing so, the useful data samples are ensured for the development of the ML model. The next step is to apply a normalisation function to all variables due to the fact that each variable has a different range of values. Thus, all variable values are scaled in the range between [0-1]. The last step is to split the data into three parts, i.e. training data, validation data and test data. Noteworthy, when developing ML models based on supervised learning, the process of labelling the data should take place before splitting the data. During this process, the fault class is assigned to the appropriate data for the classification problem. Generally, the data in each part is divided into 80\% for training, 10\% for validation and 10\% for testing part. 

\begin{figure}[thb]
    \centering
	\includegraphics[trim={0.2cm 0.2cm 0.2cm 0.2cm}, clip,width= 1\linewidth]{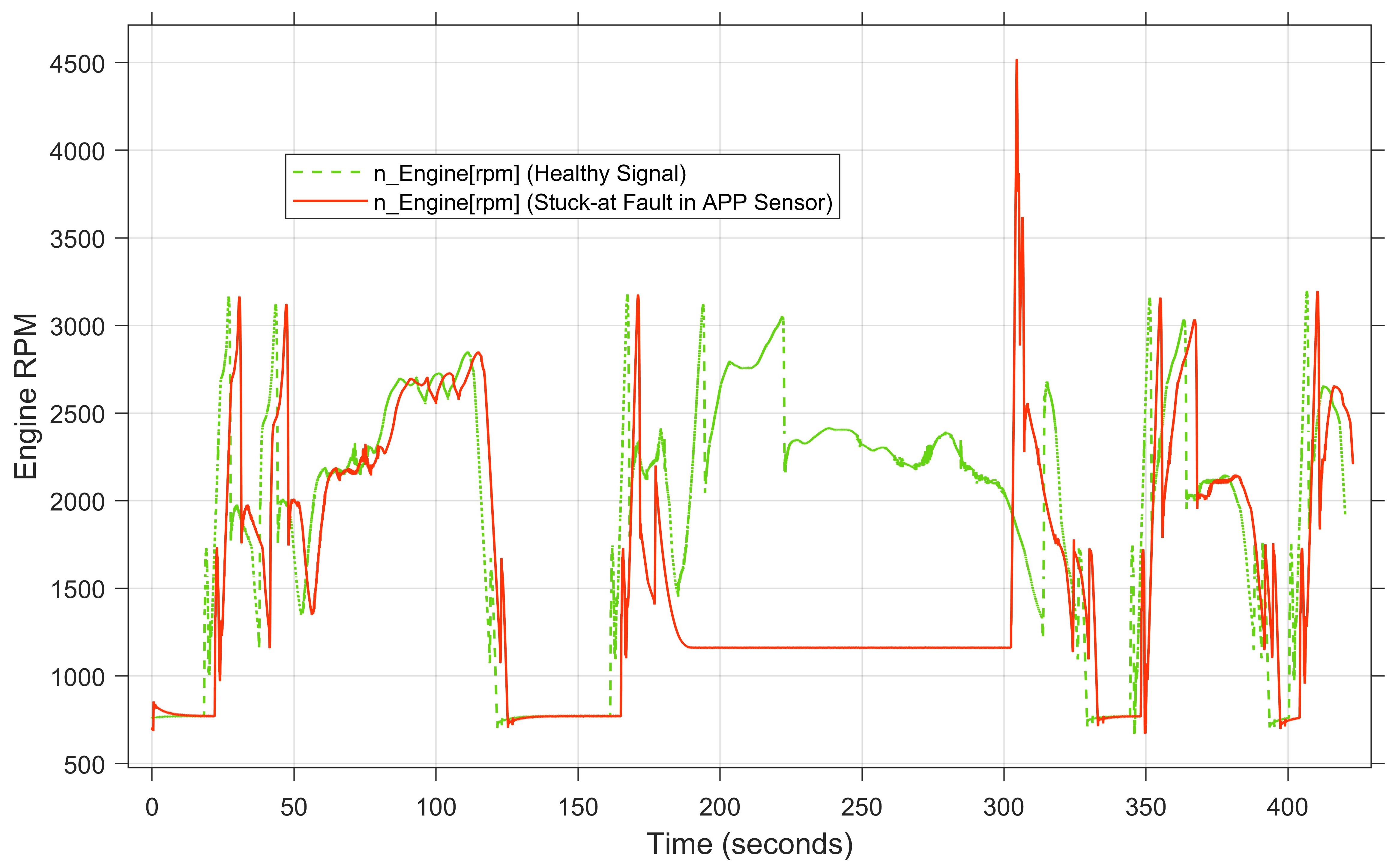}
	\caption{Engine system behaviour under stuck-at fault.}
	\label{fig: sample-figure}       
\end{figure}

\begin{figure}[thb]
    \centering
	\includegraphics[trim={0.2cm 0.2cm 0.2cm 0.2cm}, clip,width= 1\linewidth]{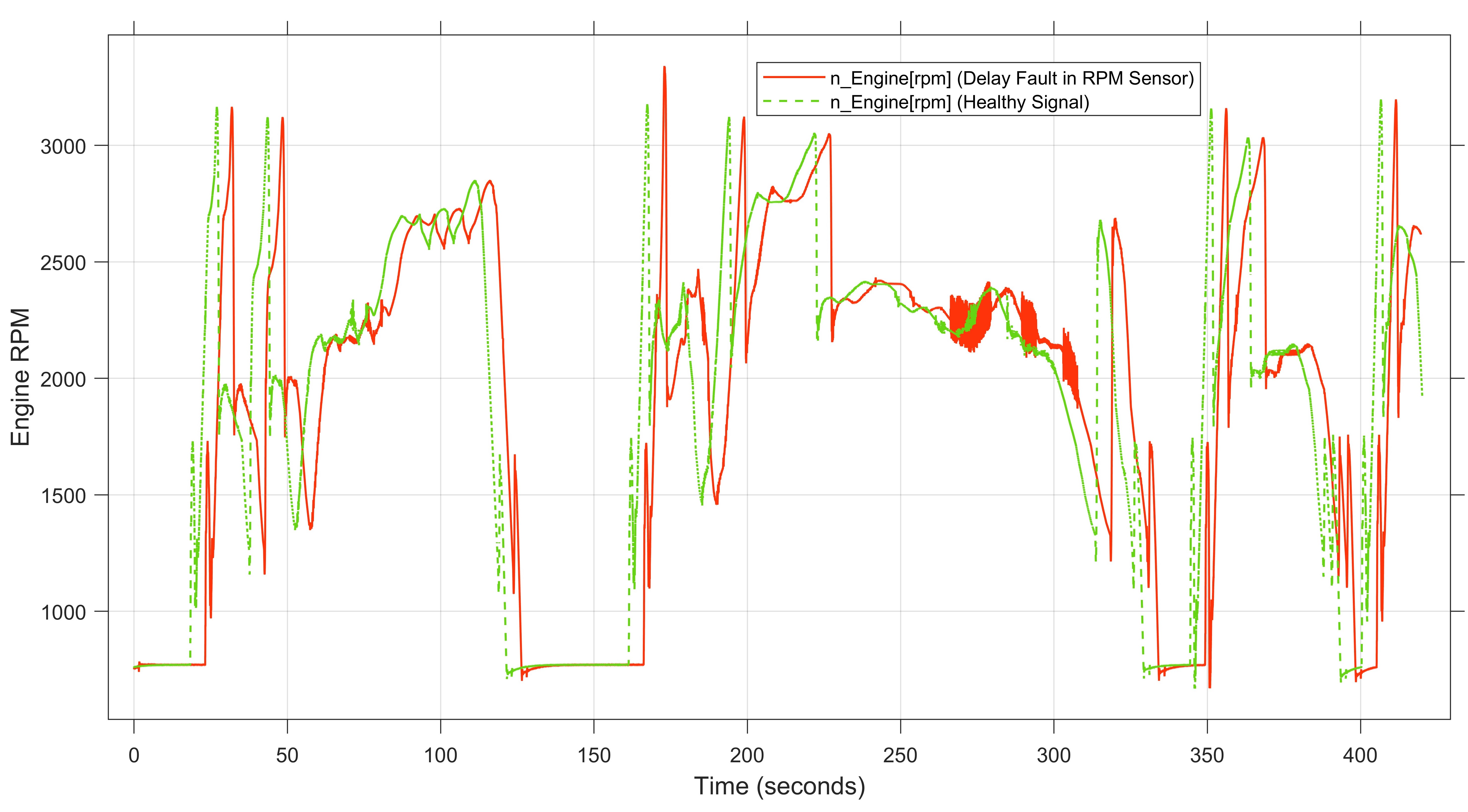}
	\caption{Engine system behaviour under delay fault.}
	\label{fig: sample-figure}       
\end{figure}

\begin{figure}[thb]
    \centering
	\includegraphics[trim={0.2cm 0.2cm 0.2cm 0.2cm}, clip,width=1\linewidth]{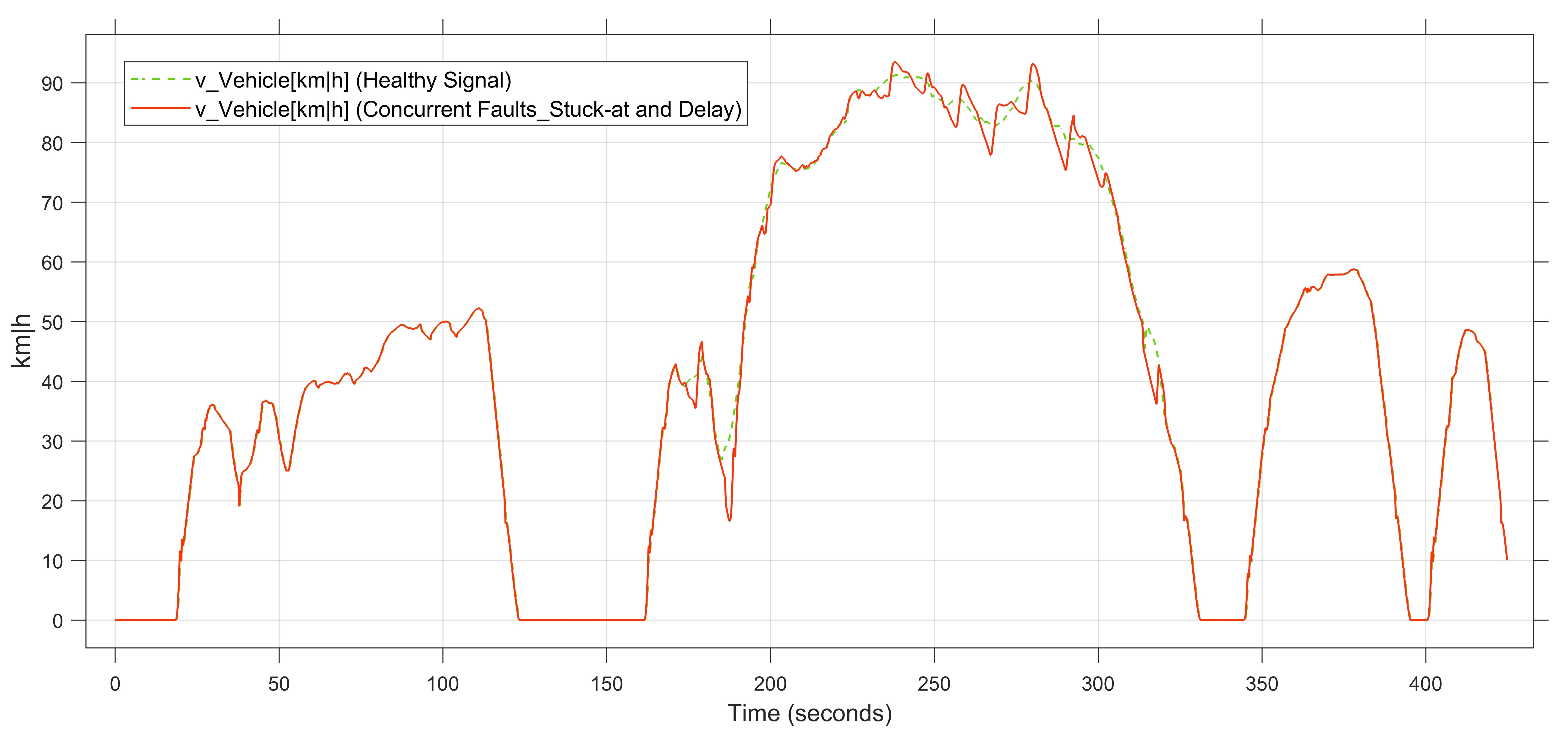}
	\caption{Vehicle system behaviour under stuck-at and delay faults simultaneously}
	\label{fig: sample-figure}       
\end{figure}

\begin{figure}[thb]
    \centering
	\includegraphics[trim={0.2cm 0.2cm 0.2cm 0.2cm}, clip,width=1\linewidth]{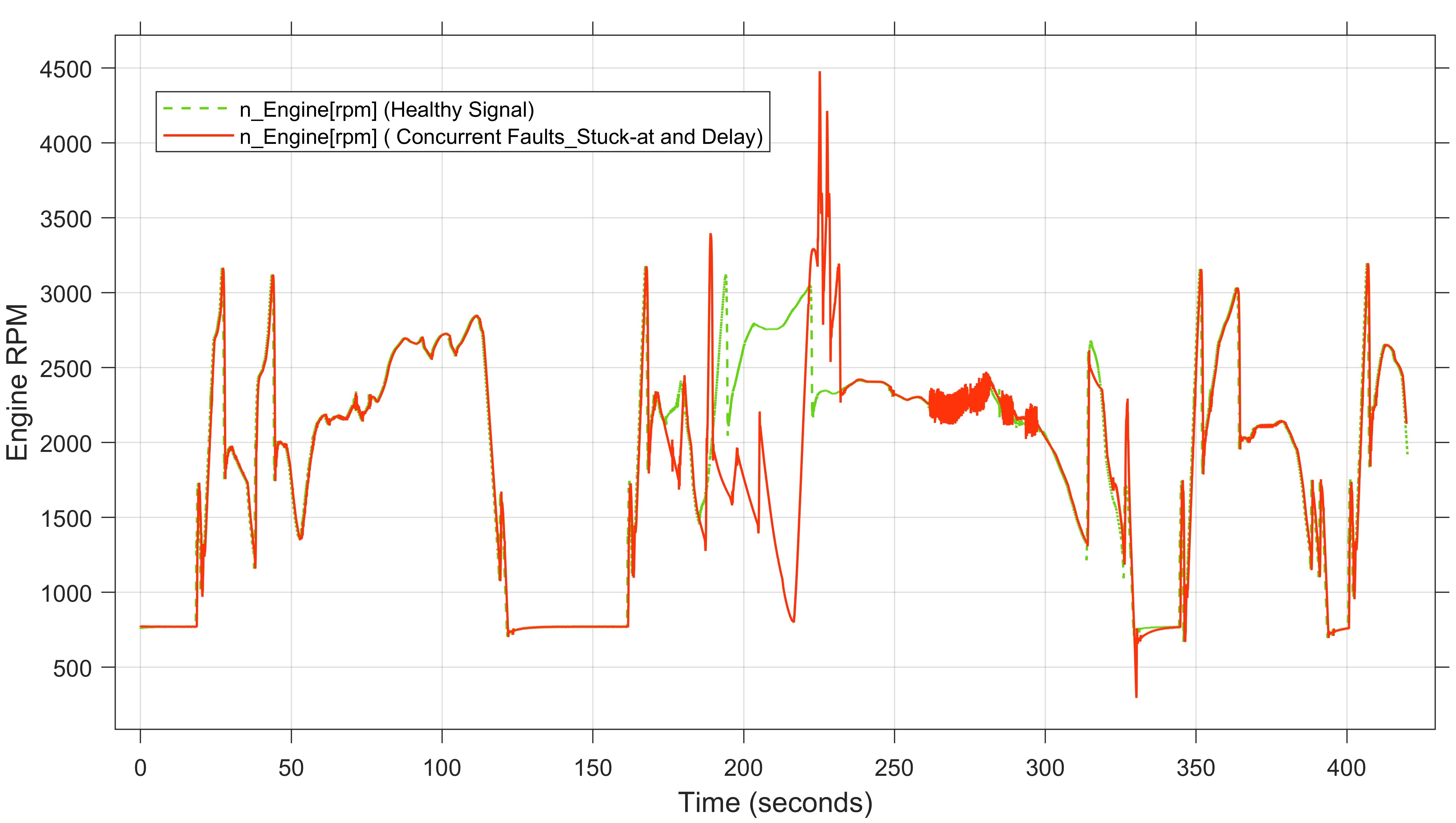}
	\caption{Engine system behaviour under stuck-at and delay faults simultaneously}
	\label{fig: sample-figure}       
\end{figure}

\begin{figure}[thb]
    \centering
	\includegraphics[trim={0.2cm 0.2cm 0.2cm 0.2cm}, clip,width=1\linewidth]{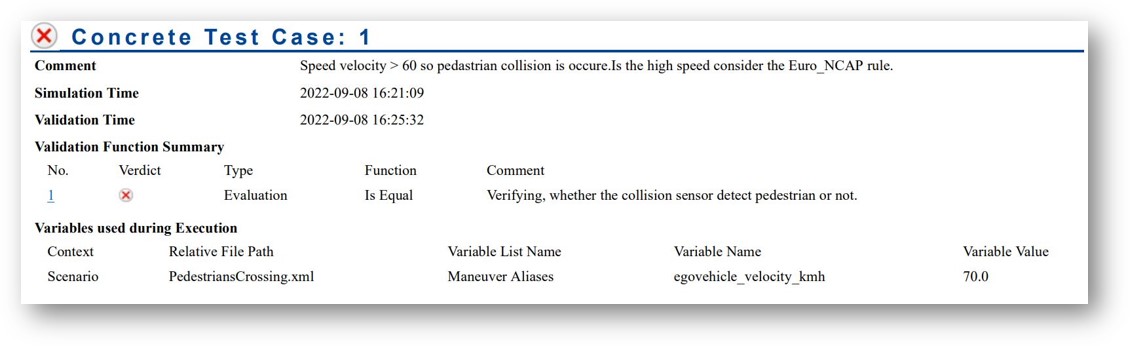}
	\caption{Generated report of test case execution}
	\label{fig: sample-figure}       
\end{figure}

As a result of the TCs execution, where the desired system signals are compared to the observed outputs, a test report is generated. What is of interest in the report is the logs, written in natural language, describing the failures that occurred. One example of the generated test report is shown in Figure 11. As can be seen, a small text is written indicating the failed TCs as information about the causes of the failure. However, the mentioned transcripts are not able to correctly identify the failure causes without the expert knowledge of the tester. Therefore, an intelligent model can be trained based on the collected logs so that the RCA process can be performed efficiently without human invention. Furthermore, the generated logs can be employed to provide additional information to support the decision of the signal-based FDD model.
Similar to the time series data, various steps are also conducted on the collected logs as a pre-processing stage before training the model. The key steps in preprocessing text data are tokenisation, normalisation, stop word removal and stemming (\cite{ahmed2022short}). After applying Natural language processing (NLP) techniques to the collected data, it is divided into three parts: training with 80\%, validation with 10\% and testing with 10\%.

\section{Conclusion and Future Work}

A novel framework for generating and collecting representative data for intelligent failure analysis applications during the system validation process is proposed in this paper. To this end, HIL simulation and fault injection method are used considering real-time constraints. The proposed framework objective is to provide the required dataset for training, testing and validation of DL-based intelligent failure analysis of HiL test records. Innovatively, there is no need for fault modelling within the system architecture in order to acquire the data. Instead, the target faults are injected programmatically in real-time without any modelling effort and time. Moreover, the proposed framework is able to simulate and collect faulty data in different formats, i.e. sequential and textual. As time-series data, the effect of hardware random faults is simulated at the system level in terms of transient and permanent faults. In this manner, the system behaviour under various critical fault conditions can be accurately and reliably captured in real-time. The higher the quality of the captured data, the higher the performance of the developed Dl model.  In this study, a complex automotive gasoline engine has been selected to validate and demonstrate the proposed approach. The outcomes show the applicability of the approach in representing the effects of single and simultaneous faults at the system level. 
Simulations of nine different types of sensor and actuator faults were carried out in this study. Besides, combination of two faults at different locations simultaneously was simulated. On the other hand, the textual descriptions of the failed TCs were generated as a result of injecting the faults in real-time during the system executions. To conclude, a high quality dataset with high faults classes coverage can be collected with low effort during real-time simulation.  

In the future, the collected data from the proposed framework will be employed to address further research problems. More specifically, the development of ML-based single and simultaneous fault detection and localisation based on supervised learning approach can be investigated. Furthermore, the proposed framework will be further improved using automation tools so that the injection process will be performed automatically, effectively and systematically according to the requirements.







\printbibliography

\end{document}

%% file: hicss-packages.tex
\usepackage[letterpaper]{geometry}
\usepackage{hicss}
\usepackage{times}
\usepackage[none]{hyphenat}
\usepackage{url}
\usepackage{latexsym}
\usepackage{minted}
\usepackage{indentfirst}
\usepackage{graphicx}
\graphicspath{{images/}}
\usepackage[
    style=apa,
  ]{biblatex}